\documentclass[twocolumn,showpacs,superscriptaddress,preprintnumbers,amsmath,amssymb,prl]{revtex4}
\usepackage{graphicx}
\usepackage{dcolumn}
\usepackage{bm}

\begin{document}

\preprint{}

\title{Time dilation of a bound half-fluxon pair \\
in a long Josephson junction with a ferromagnetic insulator
}

\author{Munehiro Nishida}
\affiliation{Graduate School of Advanced Science of Matter, 
Hiroshima University, Higashi-Hiroshima, 739-8530, Japan.
}
\author{Kyoko Murata}
\affiliation{Graduate School of Advanced Science of Matter, Hiroshima
University, Higashi-Hiroshima, 739-8530, Japan.
}
\author{Toshiyuki Fujii}
\affiliation{Graduate School of Integrated Arts and Sciences, Hiroshima
University, Higashi-Hiroshima, 739-8521, Japan.
}
\author{Noriyuki Hatakenaka}
\affiliation{Graduate School of Integrated Arts and Sciences, Hiroshima
University, Higashi-Hiroshima, 739-8521, Japan.
}

\date{ \today  }

\begin{abstract}
 The fluxon dynamics in a long Josephson junction with a ferromagnetic
 insulating layer is investigated.
 It is found that the Josephson phase obeys a double sine-Gordon equation 
 involving a bound $\pi$ fluxon solution, and the internal oscillations
 of the bound pair acting as a clock exhibit Lorentz reductions in their
 frequencies regarded as a relativistic effect in the time domain, i.e.,
 time dilation.
 This is the complement to the Lorentz contraction of fluxons with
 no clock.
 A possible observation scheme is also discussed.
\end{abstract}

\maketitle

The Lorentz transformation in space-time plays a key role in the special
theory of relativity, resulting in counterintuitive phenomena far from
our everyday experience such as length contraction when an object is
travelling close to the speed of light.
The Lorentz contraction is difficult to confirm experimentally for
energetic reasons related to macroscopic objects, i.e., macroscopic
objects cannot be accelerated to a sufficiently high velocity for the
Lorentz contraction to be observed.
Only a few examples have been demonstrated in specific systems using the
Earth's rotation \cite{Mueller,Kennedy,Hils}.
An elementary excitation with a quantum unit of magnetic flux in long
Josephson junctions, known as a fluxon, obeys sine-Gordon equations
involving the Lorentz invariance.
The length contraction of fluxons in Josephson junctions was shown in
experiments by recording the voltage pulse created by the fluxon motion
\cite{Matsuda} and by imaging the contraction of the fluxon-antifluxon
collision region in an annular Josephson junction with increasing fluxon
velocity by employing a low-temperature scanning electron microscope
\cite{Laub}.

However, a single fluxon cannot exhibit another relativistic effect in
the time domain, namely time dilation where moving clocks run slowly,
because it does not have its own clock.
As with any vibrating system, an internal degree of freedom in the
system can serve as a moving clock counting time intervals. 
One candidate is a bound pair consisting of a fluxon and an antifluxon,
and known as a breather.
The breather moves along a Josephson transmission line with the internal
oscillation in which the fluxon and the antifluxon are exchanging their
positions.
This internal oscillation acts as a clock. 
However, the breather is inadequate for testing this effect because it
is too fragile as regards external perturbations \cite{Kivshar}. 
In particular, the breather disappears even in a small dissipation because
it is topologically equivalent to the vaccum.
This feature makes it extremely difficult to observe the effect of time
dilation.
Another possibility is the $\pi$-$\pi$ kink pair solution in a
double sine-Gordon equation (DSGE) in long isolated Josephson junctions
\cite{Kasai}.
This kink pair has a stable internal oscillation below the lower edge of
the continuum phonon band \cite{Sodano}. 
Although this internal oscilation actually attenuates in the presence of
dissipation, it is possible to excite the oscillation again on
demand because the kink pair itself with a topological twist cannot
decay into vacuum.
However, it is difficult to observe 
$\pi$-$\pi$ kink oscilations in isolated systems with a non-contact
measurement.
In this Letter, we propose a new type of Josephson transmission line
that obeys a DSGE, namely a long Josephson junction with a ferromagnetic
insulator, in order to explore the time dilation of a bound pair of half
fluxons through their internal oscillations.
We also discuss an experimental scheme for observing time dilation in
such a system.

In the past decade, considerable attention has been directed towards
Josephson $\pi$ junctions characterized by a minimum Josephson coupling
energy at a phase difference of $\pi$, in relation to the physics of a
Cooper pair with a finite momentum.
The crossover between 0 and $\pi$ junction states was demonstrated in
superconductor-ferromagnet-superconductor (SFS) junctions as a function
of temperature \cite{Ryazanov_Aarts_01} and barrier thickness
\cite{Kontos_Boursier_02}. 
In the vicinity of the crossover, it is conjectured that a second order
component of Cooper-pair tunneling, i.e., $\sin(2\phi)$ with $\phi$
being the phase difference across the junction, becomes dominant in the
current-phase relation
\cite{Tanaka_Kashiwaya_97, Barash_Bobkova_02, Radovic_Vujicic_01,
Sellier_Calemczuk_04}.
This might significantly change the dynamics of the phase difference across
the junction through the current-phase relation in a long Josephson
junction.

\begin{figure}[tbph]
 \begin{center}
  \includegraphics[keepaspectratio=true,width=.34\textwidth]{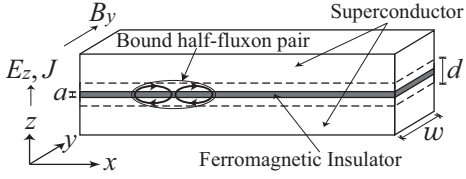}
  \caption{Schematic diagram of a long Josephson junction with a
  ferromagnetic insulator.}
  \label{fig:junction}
 \end{center}
\end{figure}

Consider two identical superconductors separated by a thin ferromagnetic
insulator with thickness $a$ lying in the $xy$ plane as shown
in Fig.\ 1.
The dimension $w$ in the $y$ direction is assumed to be much smaller
than the Josephson penetration depth, and thus the junction is
considered to be one-dimensional \cite{Laub}.
Starting as usual from Maxwell equations that takes account of the
spatiotemporal dependence of the phase difference $\phi$ across the
junction due to electric field $E_z$ and magnetic field $B_y$ through
the relations 
$E_z=(\hbar/2ea)\partial_t \phi$ and
$B_y=(\hbar/2ed)\partial_x \phi$, with $\partial_t=\partial/\partial t$
and $\partial_x=\partial/\partial x$,
we find
\begin{equation}
 \label{Maxwell}
(\hbar/2 e\mu d)\partial^2_x \phi=J+
(\hbar \epsilon/2e a)\partial^2_t\phi, 
\end{equation}
where $\epsilon$ is the dielectric constant of the ferromagnetic
insulator.
In contrast to a usual long Josephson junction, the current density 
in a long Josephson junction with a ferromagnetic insulator 
is now given as
\cite{Tanaka_Kashiwaya_97, Barash_Bobkova_02, Radovic_Vujicic_01,
Sellier_Calemczuk_04}
\begin{equation}
\label{current}
J=J_{\text{c}}\left(\lambda \sin\phi + \sin2\phi \right), 
\end{equation}
where $J_c$ is the amplitude of the current density for the second
harmonic and $\lambda$ is the ratio of the amplitude between the first
and second harmonics.
When $\lambda\simeq 0$, $J_{\text{c}}$ becomes the Josephson critical
current density.
Substituting Eq.\ (\ref{current}) into Eq.\ (\ref{Maxwell}), we obtain 
\begin{equation}
\label{DSGE}
\partial^2_x\phi  
-\partial^2_t\phi
=\lambda\sin\phi +\sin2\phi,
\end{equation}
where the coordinate $x$ is normalized by $\sqrt{2}\lambda_{\text{J}}$ where
$\lambda_{\text{J}}=\sqrt{\hbar/4e\mu d J_{\text{c}}}$ is the Josephson
penetration depth for $\lambda=0$ and the time $t$
is normalized by $\sqrt{2}\omega_{\text{J}}^{-1}$ where
$\omega_{\text{J}}=\bar{c}/\lambda_{\text{J}}$ is the frequency of the
Josephson plasma oscillation with $\bar{c}=\sqrt{a/d\epsilon\mu}$. 
The phase difference in an SFS hybrid junction thus obeys a DSGE, and
constitutes one of the main results in this paper.

A DSGE is not integrable and exact soliton solutions do not exist.
However, it is possible to find a topologically stable 2$\pi$-kink
solution for a range of parameter $\lambda$, i.e., $\lambda>0$.
The potential energy density is expressed as
$U_{\text{J}}(\phi)=\lambda(1-\cos\phi)+\frac{1}{2}(1-\cos 2\phi)$.
Here, the energy is normalized by
$E_0=2\sqrt{2}\varepsilon_{\text{J}}w\lambda_{\text{J}}$ where
$\varepsilon_{\text{J}}=\hbar J_{\text{c}}/4e$ is the specific Josephson
coupling energy per unit area for $\lambda=0$.
Since this potential has minima at $\phi=0\ (\text{mod}\ 2\pi)$ for
$\lambda>0$, there are 2$\pi$-kink solutions where the phase changes
from 0 to $\pm 2\pi$ as $x$ passes from $-\infty$ to $\infty$.
The shape of the $2\pi$ kink is determined by the competition between the
potential energy density $U_{\text{J}}(\phi)$ and the elastic energy
density, 
$U_{\text{E}}(\phi)=1/2(\partial_x \phi)^2$,
which measures the rigidity of the Josephson phase difference. 

\begin{figure}[tbph]
 \begin{center}
  \includegraphics[keepaspectratio=true,width=.34\textwidth]{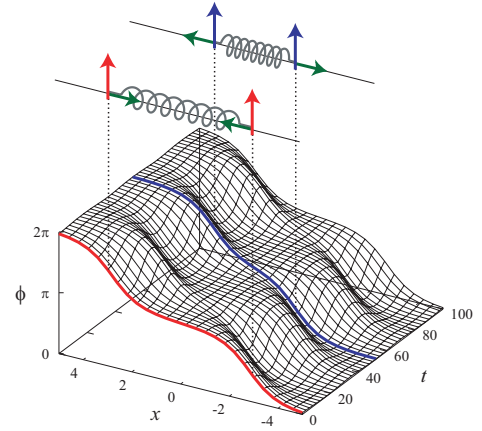}
  \caption{(Color online). Phase profile during the $\pi$-$\pi$ kink
  oscillation. Vertical (red) arrows denote the positions of $\pi$ kinks.
  Expansion and contraction of springs express the deviation from
  equilibrium separation between $\pi$ kinks symbolically and holizontal
  (green) arrows denote the directions of restoring forces.}
  \label{fig:profile}
 \end{center}
\end{figure}

For $0<\lambda<2$, the potential energy density $U_{\text{J}}$ also has
an extra local minimum at $\phi=\pi$, in addition to the $\phi=0$ minima.
This minimum leads to a new situation where the 2$\pi$ kink splits into
two separate $\pi$ kinks owing to the gain of the
Josephson energy $U_{\text{J}}$ against the elastic energy $U_{\text{E}}$.
The equilibrium separation between two $\pi$ kinks is determined by the
competition between the gain of $U_{\text{J}}$ around $\phi=\pi$ and the
loss of $U_{\text{E}}$ by the additional change in the phase difference.
With a moving $\pi$-$\pi$ kink pair, these compete in time, resulting in
a new type of dynamics. 
The $\pi$-$\pi$ kink pair exhibits an internal oscillation related to the
relative oscillations of two $\pi$ kinks around the equilibrium
separation \cite{Hudak, Majernikova_94}.
Figure\ \ref{fig:profile} shows a typical profile of phase difference
$\phi$ during the $\pi$-$\pi$ kink oscillation.

\begin{figure}[tbph]
 \begin{center}
  \includegraphics[keepaspectratio=true,width=.34\textwidth]{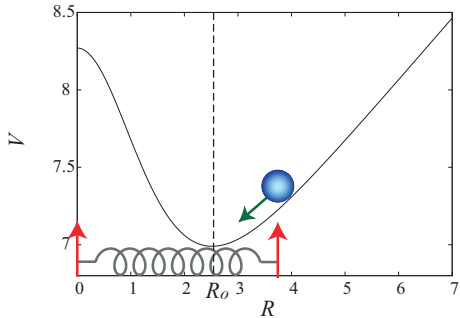}
  \caption{(Color online). Interaction potential between $\pi$
  kinks. }
  \label{fig:potential}
 \end{center}
\end{figure}

The eigenfrequency of the $\pi$-$\pi$ kink oscillation can be estimated
by using the collective coordinate method \cite{Majernikova_94}.
We assume the solution to DSGE in the form 
\begin{equation}
 \label{kink_pair}
 \phi=2\tan^{-1}\text{e}^{\Gamma\left(x+\frac{R}{2}\right)}
  +2\tan^{-1} \text{e}^{\Gamma\left(x-\frac{R}{2}\right)},
\end{equation}
where $\Gamma=\sqrt{2+\lambda}$ is the slope of each $\pi$ kink and $R$
denotes the separation between two $\pi$ kinks, which can be regarded as
a collective coordinate that describes $\pi$-$\pi$ kink oscillation.
Substituting Eq.\ (\ref{kink_pair}) into the Hamiltonian of the DSGE,
the potential energy and the inertial mass for the collective coordinate
$R$ are obtained as
{\small
\begin{align} 
 V(R)
 &= 2\Gamma
 \left[1+\tfrac{\Gamma R}{\sinh(\Gamma R)}\right]+\tfrac{4}{\Gamma}
 \left[1-\tfrac{\Gamma R}{\sinh(\Gamma R)}\right]\coth^2(\Gamma R/2)
 \nonumber \\
 &\quad +2\lambda R\coth(\Gamma R/2),\\
 M(R)&=\Gamma\left(1-\tfrac{\Gamma R}{\sinh\Gamma R}\right),
\end{align}
}\noindent
The equilibrium separation $R_0$ can be determined from the relation
$\cosh^2(\Gamma R_0/2)=1+2/\lambda$.
Thus, the $\pi$-$\pi$ kink oscillation can be regarded as the oscillation
of a particle with a mass depending on its position around the
equilibrium position $R_0$ on the anharmonic potential shown in Fig.\
\ref{fig:potential}.
In a harmonic approximation we obtain the eigenfrequency
$\omega_0(\lambda)=\sqrt{V''(R_0)/M(R_0)}$, where the prime denotes the
differentiation with respect to $R$.
This frequency of the $\pi$-$\pi$ kink oscillation serves as a clock to
measure the time. 

Since the DSGE is also Lorentz invariant, the period of the $\pi$-$\pi$
kink oscillation should increase as the velocity of the pair increases
as a consequence of the relativistic time delay
\begin{equation}
 \label{Lorentz}
 T/T_0=\left\{1-(v/\bar{c})^2\right\}^{-1/2},
\end{equation} 
where $T$ is the period of $\pi$-$\pi$ kink oscillation at the finite
velocity $v$, and $T_0$ is the period at $v=0$.
However, in an actual junction, dissipation always exists and destroys
Lorentz invariance.
Scott studied the effects of dissipation on the time dilation of a moving
breather and showed that the frequency suffers Lorentz reduction, while
the damping of oscillation is independent of velocity \cite{Scott}.
If this is also the case for $\pi$-$\pi$ kink oscillation, 
a $\pi$-$\pi$ kink pair can be used in testing the
relativistic time dilation.

To confirm the time dilation of a traveling kink pair, we performed a
numerical simulation by using a standard difference approximation with
the perturbed DSGE,
\begin{equation}
\label{eq:pDSGE}
\partial^2_t\phi
    -\partial^2_x\phi
    +\alpha\,\partial_t\phi
    +\lambda \sin\phi + \sin 2\phi = I_b,
\end{equation}
where $\alpha=1/\omega_{\text{J}}R_{\text{s}}C$ is a damping coefficient
where $R_{\text{s}}$ is
the shunt resistance and $C$ is the capacitance of the junction. 
$I_b$ is the direct bias current which keeps the velocity of the
$\pi$-$\pi$ kink pair.

\begin{figure}[tbph]
 \begin{center}
  \includegraphics[keepaspectratio=true,width=.36\textwidth]{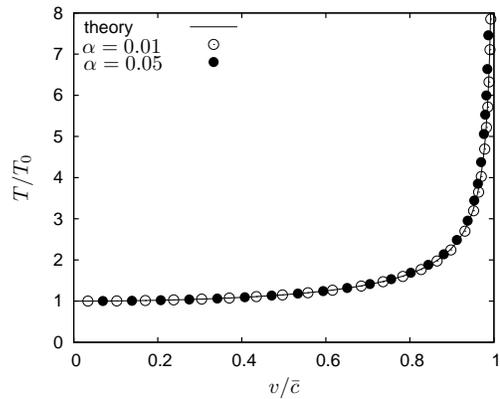}
  \caption{Velocity dependence of the $\pi$-$\pi$ kink oscillation
  period normalized by the period at $v=0$ for $\lambda=0.25$.
  The points are obtained by numerical simulations for different
  strengths of dissipation, while the solid curve is the theoretical
  expression of the time dilation, i.e., Eq.\ (\ref{Lorentz}).
  }
  \label{fig:dilation}
 \end{center}
\end{figure}

Figure \ref{fig:dilation} shows the numerical results for $T/T_0$ as a
function of the terminal velocity $v$ of a pair, which is determined
by the balance between damping ($\alpha$) and driving ($I_b$).
Here we take $\lambda=0.25$ and assume that an initial bound pair is
expressed by Eq. (\ref{kink_pair}) with appropriate separations.
The solid curve is simply the prediction obtained from the special
theory of relativity, i.e., the relativistic time delay, described by
Eq.\ (\ref{Lorentz}).
From this, we can see that our numerical results for different strengths
of dissipation are in good agreement with the theoretical predictions.
In addition, the values of $T_0$ obtained from a numerical
simulation are about 9.44 for $\alpha=0.01$ and 9.46 for $\alpha=0.05$,
which are close to the value of 9.24 estimated from $\omega_0(0.25)$.
For larger $\alpha$, we have seen the agreement with the theory in the
lower velocity region where $\alpha T \lesssim 1$, which means
underdamping.
These results indicate that the frequency of the $\pi$-$\pi$ kink
oscillation can be considered to be almost independent of $\alpha$. 
Thus, the $\pi$-$\pi$ kink oscillation is a promising candidate for use
in testing the relativistic time dilation in superconducting quantum
circuits.

Now let us discuss a possible scheme for observing the effect of the time
dilatation of a $\pi$-$\pi$ kink oscillation in a long SFS junction.
In general, it is difficult to monitor the real-time dynamics of a
$\pi$-$\pi$ kink oscillation
directly, while it might be possible to determine the eigenfrequency of
a propagating bound pair by using the resonance effect.
The bound pair exhibits a forced oscillation under an external
alternating current, leading to excitation of the $\pi$-$\pi$ kink pair. 
In particular, resonance occurs when the frequency of the external
current matches the frequency of the $\pi$-$\pi$ kink oscillation. 
With resonance, the amplitude of the oscillation becomes large, 
and thus achieve a distinguishable value compared with the equilibrium
width of the kink pair.
It may be possible to observe the $\pi$-$\pi$ kink pair by using an
existing experimental technique such as low-temperature scanning
electron microscopy \cite{Laub}, namely, by imaging the collision region
between a $\pi$-kink pair and an anti-$\pi$-kink pair in an annular SFS
Josephson junction. 
Thus, we can measure the frequency of the $\pi$-$\pi$ kink oscillation
by detecting the total width of a $\pi$-$\pi$ kink pair as a function of
the frequency of the applied oscillating current.

To study the resonance behavior, we performed a numerical simulation
based on Eq.\ (\ref{eq:pDSGE}) including external alternating
currents $I_{\text{ex}}\sin\omega t$ together with the direct bias current
$I_{\text{b}}$, where $I_{\text{ex}}$ and $\omega$ are the normalized
amplitude and frequency, respectively.
The parameters we use in the simulations are $\lambda=0.25$,
$I_{\text{ex}}=0.15$ and $\alpha=0.1$.

\begin{figure}[tbph]
 \begin{center}
  \includegraphics[keepaspectratio=true,width=.44\textwidth]{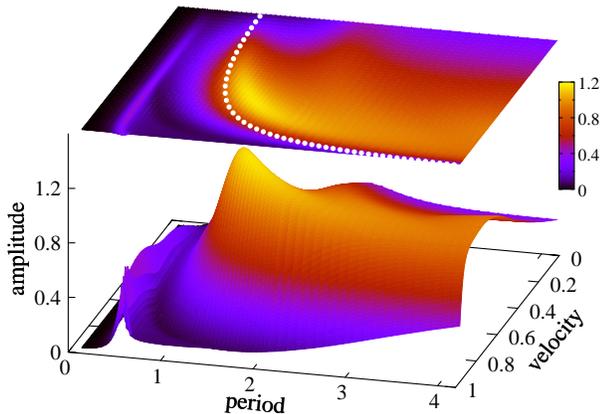}
  \caption{(Color online). Amplitude of the $\pi$-$\pi$ kink oscillation
  as a function of the period of external alternating current and
  the velocity of the bound fluxon pair.
  The (white) dotted line denotes the relation of the relativistic time
  dilation Eq.\ (\ref{Lorentz}).}
  \label{fig:v_t_a}
 \end{center}
\end{figure}

Figure \ref{fig:v_t_a} shows the amplitude of the oscillation of the
separation between $\pi$ kinks as a function of the normalized period of
external alternating current $2\pi/\omega T_0$ with $T_0=9.54$ and the
terminal velocity of a pair $v/\bar{c}$.
The amplitude is estimated from the half value of the difference between
the maximum and minimum separation of the kink pair, and is normalized
by the width of a moving $\pi$ kink with velocity $v$, i.e.,
$\sqrt{2}\lambda_{\text{J}}\sqrt{1-(v/\bar{c})^2}/\Gamma$.
We can easily find the resonance of the $\pi$-$\pi$ kink oscillation as
indicated by a (white) dotted line.
This line shows the Lorentz dilation expressed by Eq.\
(\ref{Lorentz}).
In addition, there is another resonance with $1/2$ harmonic that
has a period twice that of eigen oscillation as a result of a nonlinear
effect, called nonlinear resonance \cite{Landau}.
As seen in Fig.\ \ref{fig:potential}, a nonlinear interaction is induced
between $\pi$ kinks when the amplitude of the $\pi$-$\pi$ kink
oscillation becomes large because of an increase in the external field
amplitude.
In contrast to a linear interaction yielding a harmonic resonance, the
nonlinear interaction gives rise to a subharmonic resonance at a double
eigenfrequency.
The $1/2$-harmonic resonance corresponds to the cubic nonlinearity of
the asymmetric potential shown in Fig.\ \ref{fig:potential}.
There is also another resonance that is almost independent of the pair
velocity around a period of $0.5$.
This resonance corresponds to a homogeneous Josephson plasma oscillation
whose period $T_{\text{JP}}$ is determined by the curvature of
$U_{\text{J}}$ around $\phi=0$ minima, i.e.,
$T_{\text{JP}}/T_0=2\pi/\sqrt{U''_{\text{J}}(0)}T_0
=2\pi/\sqrt{\lambda+2}T_0\simeq 0.44$, 
where the prime denotes the differentiation with respect to $\phi$.

Finally, let us roughly estimate the parameters.
In the vicinity of the 0-$\pi$ crossover, Josephson critical current
would be quite small, say, two orders of magnitude smaller than that of
usual junction \cite{Barash_Bobkova_02, Sellier_Calemczuk_04}, and thus
$\lambda_{\text{J}}\sim$100 $\mu$m, $\omega_{\text{J}}\sim$10 GHz and
$\alpha\sim 0.1$.
The amplitude of $\pi$-$\pi$ kink oscillation at the resonance is of the
order of 10-100 $\mu$m,
which is within the range of spatial resolution of the experiment
\cite{Laub}.

In summary, we have studied fluxon dynamics in a long Josephson junction
with a ferromagnetic insulator.
In such a system, the Josephson current-phase relation differs from 
that of a conventional junction, resulting in half-integer fluxons 
that obey a double sine-Gordon equation. 
The bound pair of the half-fluxons exhibits an internal oscillation that
is unique to this system, and could be detected through this oscillation
by using the resonance effect. 
This might provide evidence for the existence of a second harmonic
component in the current-phase relation of an SFS hybrid junction. 
In addition, we demonstrated numerically the time dilation of the bound
half-fluxon pair by detecting the Lorentz reduction in its frequency as
a function of pair velocity.
This is the complement to the Lorentz contraction. 
Moreover, a quantized $\pi-\pi$ kink pair might also provide 
an application for Josephson-based quantum computers as a new type of
{\it mobile} qubit using Josephson $\pi$ states \cite{Yamashita}.

\begin{acknowledgments}
This work was supported in part by KAKENHI (17740267 and 18540352) from
 MEXT of Japan, and by a Grant-in-Aid for JSPS Fellows (195836).
\end{acknowledgments}


\begin{thebibliography}{99}
 \bibitem{Mueller}
	 J. Mueller and M. Soffel, 
	 Phys. Lett. A \textbf{198}, 71 (1995). 
 \bibitem{Kennedy}
	 R. J. Kennedy and E. M. Thorndike, 
	 Phys. Rev. \textbf{42}, 400 (1932). 
 \bibitem{Hils}
	 D. Hils and J. L. Hall, 
	 Phys. Rev. Lett. \textbf{64}, 1697 (1990). 
 \bibitem{Matsuda}
	 A. Matsuda and T. Kawakami, 
	 Phys. Rev. Lett. \textbf{51}, 694 (1983). 
 \bibitem{Laub}
	 A. Laub, T. Doderer, S. G. Lachenmann, R. P. Huebener, and
	 V. A. Oboznov,
	 Phys. Rev. Lett. \textbf{75}, 1372 (1995). 
 \bibitem{Kivshar}
     Yu. S. Kivshar and B. A. Malomed, 
     Rev. Mod. Phys. \textbf{61}, 763 (1989). 
 \bibitem{Kasai}
	 Y. Kasai, S. Tanda, N. Hatakenaka, and H. Takayanagi, 
	 Physica C \textbf{352}, 211 (2001). 
 \bibitem{Sodano}
	 P. Sodano, M. El-Batanouny, and C. R. Willis,
	 Phys. Rev. B \textbf{34}, 4936 (1986).
 \bibitem{Ryazanov_Aarts_01}
	 V. V. Ryazanov, V. A. Oboznov, A. Y. Rusanov,
	 A. V. Veretennikov, A. A. Golubov, and J. Aarts, Phys. Rev.
	 Lett. \textbf{86}, 2427 (2001).
 \bibitem{Kontos_Boursier_02}
	 T. Kontos, M. Aprili, J. Lesueur, F. Genet, B. Stephanidis, and
	 R. Boursier, Phys. Rev. Lett. \textbf{89}, 137007 (2002).
 \bibitem{Tanaka_Kashiwaya_97}
	 Y. Tanaka and S. Kashiwaya, Physica C \textbf{274}, 357 (1997).
 \bibitem{Barash_Bobkova_02}
	 Yu. S. Barash and I. V. Bobkova,
	 Phys. Rev. B \textbf{65}, 144502 (2002). 
 \bibitem{Radovic_Vujicic_01}
	 Z. Radovi\'{c}, L. Dobrosavljevi\'{c}-Gruji\'{c}, and
	 B. Vuji\v{c}i\'{c}, Phys. Rev. B \textbf{63}, 214512 (2001).
 \bibitem{Sellier_Calemczuk_04}
	 H. Sellier, C. Baraduc, F. Lefloch, and R. Calemczuk, Phys.
	 Rev. Lett. \textbf{92}, 257005 (2004).
 \bibitem{Hudak}
	 O. Hud\'{a}k, Phys. Lett. A \textbf{86}, 208 (1981). 
 \bibitem{Majernikova_94}
	 E. Majern\'{i}kov\'{a}, Phys. Rev. E \textbf{49}, 3360 (1994).
 \bibitem{Scott}
	 A. C. Scott,
	 Phys. Scr. \textbf{20}, 509 (1979).
 \bibitem{Landau}
	L. D. Landau and E. M. Lifshitz, 
	\textit{Mechanics (Third Edition),
	Course of Theoretical Physics Volume 1} 
	(Elsevier Butterworth-Heinemann, 1976)
 \bibitem{Yamashita}
	 T. Yamashita, K. Tanikawa, S. Takahashi, and S. Maekawa, 
	 Phys. Rev. Lett. \textbf{95}, 097001 (2005). 
\end{thebibliography}
\end{document}